\documentclass[%
aps,
  pra,%
 amsmath,amssymb,
  reprint,%
 floatfix,
onecolumn,
notitlepage
]{revtex4-1}

\usepackage{graphicx}% Include figure files
\usepackage{dcolumn}% Align table columns on decimal point
\usepackage{bm}% bold math

\allowdisplaybreaks[2]

\newcommand{\PR }{{\it Phys. Rev.} }
\newcommand{\PRL }{{\it Phys. Rev. Lett.} }
\newcommand{\PRA }{{\it \PR A} }

\newcommand{\Sc }{{\it Science} }
\newcommand{\Nat }{{\it Nature} }
\newcommand{\PLA }{{\it Phys. Lett. A} }
\newcommand{\vv}[1]{\mathbf{#1}}         %%% vector symbols (std)
\newcommand{\mathperiod}{\,.} %%% end of sentence in mathematical environment
\newcommand{\mathcomma}{\,,}  %%% comma in mathematical environment
\newcommand{\ee}{\mathrm{e}}             %%% exponential function
\newcommand{\ii}{\mathrm{i}}             %%% the imaginary unit i^2=-1
\newcommand{\abs}[1]{\left|{#1}\right|}
\newcommand{\abssmall}[1]{|{#1}|}

\newcommand{\ketsmall}[1]{|{#1}\rangle}

\newcommand{\ie}{\mbox{i.\,e.}\nolinebreak[4]}
\newcommand{\eg}{\mbox{e.\,g.}\nolinebreak[4]}

\newcommand{\anfz}[1]{``{#1}''}

\begin{document}

\title{Tailoring superradiance to design artificial quantum systems}

\author{Paolo~Longo}
\email{paolo.longo@mpi-hd.mpg.de}
\author{Christoph H. Keitel}%
\author{J\"org~Evers}%
\email{joerg.evers@mpi-hd.mpg.de}
\affiliation{Max Planck Institute for Nuclear Physics, Saupfercheckweg 1, 69117 Heidelberg, Germany}

\date{\today}

\begin{abstract}
Cooperative phenomena arising due to the coupling of individual atoms via the radiation field are a cornerstone of modern quantum and optical physics. 
Recent experiments on x-ray quantum optics added a new twist to this line of research 
by exploiting superradiance in order to construct artificial quantum systems. 
However, so far, systematic approaches to deliberately design superradiance properties are lacking,  
impeding the desired implementation of more advanced quantum optical schemes. 
Here, we develop an analytical framework for the engineering of single-photon superradiance in extended media
applicable across the entire electromagnetic spectrum, 
and show how it can be used to tailor the properties of an artificial quantum system.
This \anfz{reverse engineering} of superradiance not only provides an avenue towards non-linear and quantum mechanical phenomena at x-ray energies, 
but also leads to a unified view on and a better understanding of superradiance 
across different physical systems.
\end{abstract}

\maketitle

A single atom coupled to an environment is usually subject to spontaneous emission and 
experiences a frequency shift referred to as the Lamb shift.
In an aggregation of atoms coupled via the radiation field, collective effects 
can significantly alter the properties compared to a single emitter. 
For instance, this was realised by Dicke \cite{dicke54,haroche82}, who showed that~$N$ identical atoms
confined to a volume much smaller than a wavelength cubed
collectively behave as one \anfz{super atom}.
This leads to exaggerated properties such as an acceleration of spontaneous decay  by a factor of
$\chi_{\mathrm{Dicke}}=N$ (known as superradiance)
or an enhanced frequency shift (sometimes also termed \anfz{collective Lamb shift}). 
Recently, also the correlated emission from \emph{extended} ensembles of emitters
has become the focus of experimental and theoretical \cite{manassah73,manassah09,scully09}
investigations, where either the system size and/or the minimal interatomic distance~$a$ exceeds the scale of the characteristic wavelength $\lambda_0$. 
The systems considered cover a wide range of possible realisations,
including 
atoms near a nanofiber~\cite{hakuta05},
thin vapor layers~\cite{adams12},
cold atomic ensembles~\cite{browaeys,ozeri,kimble15,kaiser,javanainen14}, 
or thin-film cavities with embedded M\"ossbauer nuclei in the realm of 
x-ray quantum optics~\cite{roehlsb10,roehlsb12,heegprl,heegslow,heegfano,heeg13,kocharovskaya14,xrayqo}.

The present work is motivated by the observation that in particular the latter experiments in the field of 
nuclear quantum optics exploited a deliberate control of superradiance properties, going beyond a mere characterisation.
For instance, the observation of electromagnetically induced transparency at x-ray frequencies~\cite{roehlsb12} was enabled by the engineering of two distinct ensembles with
different superradiance properties in a single sample. Another example is the implementation of spontaneously generated coherences~\cite{heegprl}, which relied on the realization of a spatially anisotropic electromagnetic environment via superradiance. 
In both cases, superradiance was employed to design an artificial quantum system, which in turn enabled the observation of the desired effect. 

This raises the question whether a systematic and constructive approach could be established to exploit superradiance for the design of artificial quantum systems. 
Such design capabilities could overcome the limited resources accessible in state-of-the-art experiments, and thereby enable more advanced level schemes required, e.g., for the exploration of non-linear and quantum effects at x-ray energies. 

Here, we address this question by developing an analytical framework for superradiance in extended media encompassing different system dimensionalities, interatomic couplings, and environments. As our main result, we then derive expressions describing how collective decay rates and frequency shifts can be controlled in extended media, and show how they can be used for the design of an artificial optical transition.

We start with a single two-level atom
(bare transition frequency $\omega_0=c k_0$, $k_0=2\pi/\lambda_0$, $c$ is the speed of light)
which is embedded in an electromagnetic environment (\eg, free space) 
and is characterised by its spontaneous decay rate $\mathrm{Re}(V_0) \equiv \gamma_0$
(assuming Markovian reservoirs~\cite{ficek}).
The coupling to the environment also results in a frequency shift
$\mathrm{Im}(V_0)/2 \equiv \delta\omega_0$ (single-atom Lamb shift).
In the presence of an identical, second atom,
photons can be exchanged between the two atoms. Due to irreversible loss to the reservoir,
the inter-atomic coupling 
$V_{\vv{r}_i \vv{r}_j} = \gamma_{\vv{r_i} \vv{r_j}} + 2\ii \delta\omega_{\vv{r_i} \vv{r_j}}$
is complex~\cite{li12,longo14prl,longo14pra,ficek,agarwal}.
Here, $\gamma_{\vv{r_i} \vv{r_j}}$ ($\delta\omega_{\vv{r_i} \vv{r_j}}$)
represents the real-valued cross-damping (cross-coupling) term
for two atoms located at positions $\vv{r}_i$ and $\vv{r}_j$, respectively.
Considering all pair-wise couplings in an ensemble of $N \gg 1$ atoms, we find~\cite{longo14pra,li12}
\begin{equation}
 \label{eq:eigenproblem}
 0 = - \frac{\ii}{2} \sum_{j=1}^N V_{\vv{r}_i \vv{r}_j} \varphi_{\vv{r}_j}
 - (E - \omega_0) \varphi_{\vv{r}_i}
 \mathcomma
\end{equation}
where $E$ denotes the complex eigenenergy of
the collective single-excitation atomic state 
$\ketsmall{\Psi} = \sum_{i=1}^N \varphi_{\vv{r}_i} \sigma^+_{i} \ketsmall{0}$
($\sigma^+_{i}$ and $\ketsmall{0}$ signify 
the atomic raising operator for atom~$i$ and the vacuum state, respectively).
Equation~(\ref{eq:eigenproblem}) is valid for all dimensions $d$ of the atomic 
arrangement and for all (physically reasonable) couplings  $V_{\vv{r}_i \vv{r}_j}$.
Collective decay rates and frequency shifts are obtained via 
$\Gamma \equiv -2\mathrm{Im}(E)$ and $\Delta \equiv \mathrm{Re}(E) - \omega_0$, 
respectively~\cite{li12,longo14pra}.

\begin{figure}[t]
 \centering
 \includegraphics[width=\textwidth]{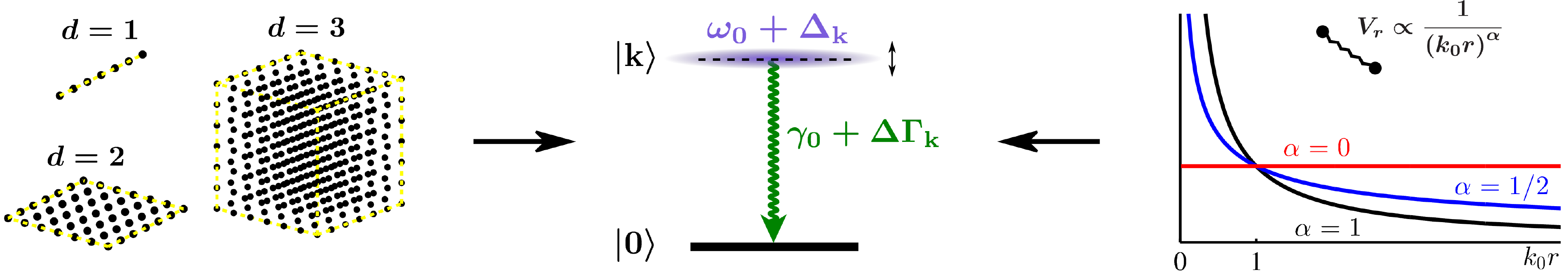}
\caption{\label{fig:dimexamples}
	  {\bf Design of an artificial optical transition through tailored superradiance.} 
	  A $d$-dimensional lattice of atoms is embedded 
	  into an electromagnetic reservoir that mediates 
	  an inter-atomic coupling $V_r \propto 1/r^\alpha$,
	  where atoms are separated by a distance $r$
	  and the coefficient $\alpha$ characterises the distance-dependence
	  (see eq.~(\ref{eq:gencoupling})).
	  We show that the resulting collective eigenstates can be utilised for
	  the implementation of an artificial transition with tunable decay rate
	  and transition frequency.
	 }
\end{figure}

In Dicke's small-volume limit, all atoms couple to each other with equal strength,
leading to a collective decay rate 
$\Gamma =N \gamma_0=\chi_{\mathrm{Dicke}} \gamma_0$
and a frequency shift
$\Delta = \chi_{\mathrm{Dicke}} \delta\omega_0$ with an enhancement factor $\chi_{\mathrm{Dicke}}$
(see methods).
To describe an extended sample, 
we consider ordered atomic arrangements, and focus on chains ($d=1$), square lattices ($d=2$),
and simple cubic lattices ($d=3$), see fig.~\ref{fig:dimexamples}.
The smallest inter-atomic distance is given by the lattice constant $a$.
Such ordered arrays are naturally provided by crystalline samples
(\eg, solid state targets employed in x-ray quantum optics~\cite{roehlsb10,roehlsb12,heegprl,heegslow,heegfano,heeg13},
optical lattices of atoms~\cite{jacksch05},
or atom--cavity networks~\cite{kay08}).
Furthermore, 
we consider a generic class of inter-atomic couplings
\begin{equation}
 \label{eq:gencoupling}
 \frac{V_r}{\gamma_0} = A_d \sin^2\theta \frac{\ee^{\epsilon \ii k_0 r} }{\left(k_0 r\right)^\alpha}~~~~~(\alpha \geq 0,~\epsilon=\pm)
 \mathcomma
\end{equation}
which depend on the distance $r$ between atom pairs.
Here, the coefficient $\alpha$ classifies the distance-dependence
and $A_d$ is a dimensionless coupling strength.
We also assume the atomic dipole moments to be 
uniformly aligned along the $x_3$ axis.
This orientation dependence is taken into account 
by the angle~$\theta$ (see methods for further details).
Since multiple terms of type (\ref{eq:gencoupling}) can be accounted 
for by a linear combination, in particular also the three common implementations of three-dimensional free space~\cite{li12}, 
atoms confined to two spatial dimensions~\cite{couto13}, 
or atoms coupled to a one-dimensional waveguide~\cite{vidal11} are covered. 
The coupling parameters for these three examples are specified in table~\ref{tab:dimexpr}.
Note that in principle $\alpha$ can be artificially engineered and controlled
as has recently been demonstrated at optical frequencies~\cite{kimble15}.

\begin{table}[b]
 \centering
\begin{tabular}{|c|c|c|c|c|c|c|}
 \hline
 $d$ 	&  $\mathrm{Re}[A_d]$	& $\mathrm{Im}[A_d]$	& $b_d$				& $c_d$ 					& $g_d( \cdot )$	\\
 \hline
 $1$ 	& $\geq 0$		& $=0$			& $\frac{1}{2}$			& $1$ 					& $\cos(\cdot)$		 \\
 \hline
 $2$	& $\geq 0$		& $=0$			&  $\frac{1}{\sqrt{\pi}}$ 	& $\sqrt{2 \pi}$				& $\cos(\cdot)$	\\
 \hline
 $3$ 	& $= 0$			& $\leq 0$		&  $\sqrt[3]{\frac{3}{4\pi}}$ 	& $2\pi \cdot \sin^2\vartheta$ 		& $\sin(\cdot)$ \\
 \hline
\end{tabular}
\caption{\label{tab:dimexpr}
	  {\bf Dimension-dependent quantities.} 
	  The table summarises the quantities appearing in eqs.~(\ref{eq:enhfac}), 
	  (\ref{eq:radialint}),
	  (\ref{eq:radialintJ}),
	  and~(\ref{eq:chiformuls1})-(\ref{eq:chiformuls4}) as function of the system dimension $d$
	  for the three considered example cases.
	  Here, $\vartheta=\arccos(\vv{k} \hat{\vv{e}}_{x_3} / k)$ 
	  denotes the angle between the eigenstate's wavevector
	  $\vv{k}$ and the $x_3$ axis.
	 }
\end{table}

\section*{Results and discussion}
The solution of eigenproblem (\ref{eq:eigenproblem}) (see methods)
reveals that those eigenstates $\ketsmall{\vv{k}}$ whose wavevector's magnitude matches the wavenumber
set by the single atom transition, \ie, $k=\abssmall{\vv{k}}=k_0$, 
exhibit the maximum possible decay rate~$\Gamma_{\mathrm{max}}=\chi_{\mathrm{max}} \gamma_0$
if the constraint 
\begin{equation}
 \label{eq:constraint}
 0 \leq \alpha < \frac{d+1}{2} 
\end{equation}
is fulfilled.
This criterion is a necessary condition for the emergence of 
superradiance and represents bounds 
on the allowed power laws of the coupling terms (exponent $\alpha$ in eq.~(\ref{eq:gencoupling}))
as a function of the lattice dimension $d$.
For the remainder, we assume that eq.~(\ref{eq:constraint}) is satisfied.
The enhancement factor~$\chi_{\mathrm{max}}$ is 
(see table~\ref{tab:dimexpr} for quantities $b_d$ and $c_d$, and methods for the prefactor ${L}_d(\alpha)$)
\begin{align}
 \chi_{\mathrm{max}} \equiv
 \chi_{k=k_0} &= 1 + \frac{ \abs{A_d} c_d \left( b_d \right)^{\frac{d+1}{2}-\alpha}  }{\frac{d+1}{2}-\alpha}  \left( k_0 a \right)^{ \frac{1-d}{2} - \alpha } \sqrt[d]{N}^{\frac{d+1}{2}-\alpha}
 \label{eq:enhfac}
\\
 &= 1 + {L}_d(\alpha) \cdot
  \left( \frac{\lambda_0}{\sqrt[d]{\mathcal{V}}} \right)^{\frac{1}{2}(d-1)+\alpha} \cdot N
 \mathperiod \label{eq:chiformula}
\end{align}
In contrast to the maximum collective decay rate, 
we find that  the collective frequency shift at $k=k_0$ is always zero 
independent of the actual physical realisation.
We thus conclude that the case of maximum superradiance is unsuitable for a control of both 
collective decay rates and frequency shifts.

\begin{figure}[t]
 \centering
 \includegraphics[width=0.48\textwidth]{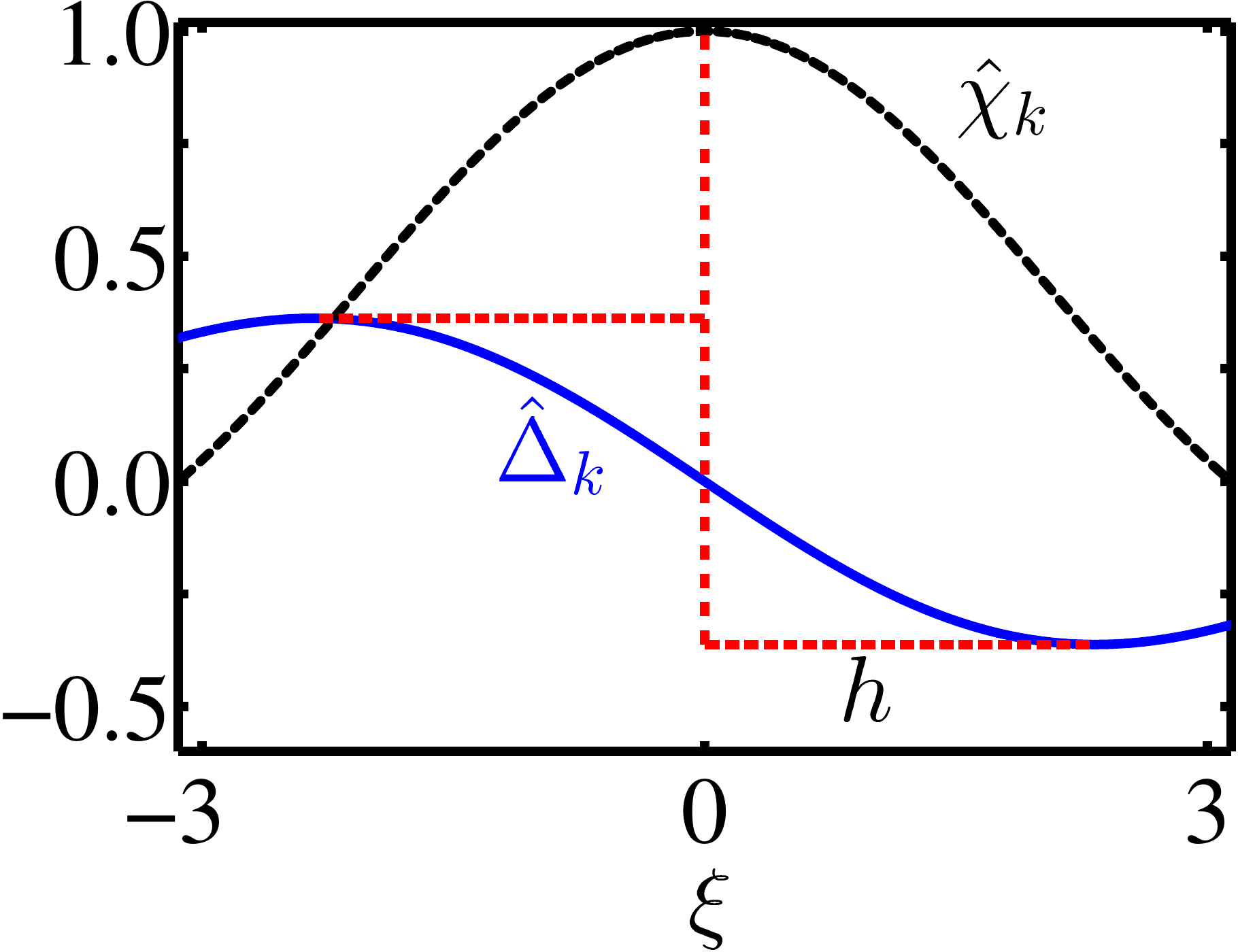}
 \caption{\label{fig:intbeta0}
	  {\bf Collective decay rates and frequency shifts.}
	  Decay rates  (black dashed curve) and frequency shifts (blue solid curve) as function of the wavenumber for $\alpha=(d-1)/2$. 
	  The figure is valid independent of dimensionality and coupling type, due to the  scaling of decay rate $\hat{\chi}_k \equiv (\chi_k-1)/(\chi_{\mathrm{max}}-1)$, shift $\hat{\Delta}_k \equiv[(\Delta_k-\delta\omega_0)/\gamma_0]/[\epsilon (\chi_{\mathrm{max}}-1)]$ and wavenumber $\xi \equiv (k-k_0)a b_d \sqrt[d]{N}$. 
	Note the offset~$h$ between the extrema
	  of the frequency shift and the decay rate maximum.
	 }
\end{figure}

To circumvent this problem, we also consider states with wavenumbers around $k = k_0$. 
Indeed, for a large but finite system,
also states with a wavenumber close to $k_0$ can exhibit an enhanced decay rate. 
We illustrate this for the most relevant case $\alpha=(d-1)/2$ 
(which includes the three common implementations mentioned below eq.~(\ref{eq:gencoupling})).
For $\abssmall{k-k_0}a \ll 1$,
we find 
\begin{eqnarray}
 \label{eq:ratesfiniteM}
 && \frac{\Gamma_{{k}}}{\gamma_0} =
 \chi_{{k}}
 \simeq
 1+
 \left(\chi_{\mathrm{max}} - 1 \right) \cdot 
 \mathrm{sinc}(\xi)
 \mathcomma \\
 \label{eq:shiftsfiniteM}
 && \frac{  \Delta_{{k}} - \delta\omega_0  }{\gamma_0} \simeq
 \epsilon \cdot
 \frac{\chi_{\mathrm{max}} - 1}{2} 
 \cdot
 \frac{\cos(\xi) - 1}{\xi}
 \mathcomma
\end{eqnarray}
where $\xi \equiv (k-k_0)a b_d \sqrt[d]{N}$ and $\mathrm{sinc}(\xi)\equiv\sin(\xi)/\xi$.
Results are shown in fig.~\ref{fig:intbeta0}, scaled in such a way that they encompass different dimensions and coupling types. 
As mentioned before, those states which are maximally superradiant at $k=k_0$ do not exhibit
a collective frequency shift.
Rather, the frequency shift's first two extrema around $k_0$ occur at wavenumbers
$k_{\pm} \equiv k_0 \pm h/(a b_d \sqrt[d]{N})$
(where $h \simeq 2.3311$).
This finding represents a unique feature as it is independent of the actual realisation 
and provides a signature suitable for a direct experimental test.

Equations~(\ref{eq:ratesfiniteM}) and~(\ref{eq:shiftsfiniteM}) also offer means to design an artificial optical transition with desired decay rate and frequency shift. 
In fact, the enhancement factor~$\chi_{\mathrm{max}}$ represents a characteristic scale for both decay rates and frequency shifts. 
As expected, we find that the particle number~$N$ and/or the sample volume~$\mathcal{V}$ can be used to control~$\chi_{\mathrm{max}}$. 
But additionally, eqs.~(\ref{eq:enhfac}) and (\ref{eq:chiformula}) explain how the dimensionality~$d$, 
the type of the inter-atomic coupling as described by~$\alpha$, 
as well as the coupling strength to the environment can be used to manipulate the  enhancement factor. 
This is of particular relevance, since these parameters could also be tuned {\it in situ} \cite{kimble15,kimble14}.
However, as mentioned previously, these quantities are not sufficient to change the ratio between decay rate and frequency shift. 
This only becomes possible by also controlling the wave number $k$ (see fig.~\ref{fig:intbeta0}).
Experimentally, the wavenumber could be adjusted via the excitation angle of the probing light field.

From a broader perspective, our results also enable us to understand how superradiant states from different realisations can be compared and categorised. 
This is important, \eg, if superradiant ensembles realised using different individual constituents are to be combined to an effective artificial quantum system. 
To this end, suppose that we can control the atom number and the volume such that 
$N \rightarrow \tilde{N} \equiv f_{N} N$ and $\mathcal{V} \rightarrow \tilde{\mathcal{V}} \equiv f_{\mathcal{V}} \mathcal{V}$, respectively,
where $f_{N}$ and $f_{\mathcal{V}}$ are arbitrary positive real numbers.
Under this transformation, the enhancement factor changes as
\begin{equation}
 \label{eq:chitransform}
 \chi_{\mathrm{max}} \rightarrow \tilde{\chi}_{\mathrm{max}} = f_{N}  \sqrt[d]{f_{\mathcal{V}}}^{\frac{1}{2}(1-d-2\alpha)} \chi_{\mathrm{max}}
 \mathperiod
\end{equation}
This behaviour allows us to classify superradiant states 
from systems with different dimensionality and types of coupling.
For instance, we may say that two extended samples characterised by $(d, \alpha)$ and $(d^\prime, \alpha^\prime)$,
respectively, are similar if they satisfy the same transformation rule (\ref{eq:chitransform})
(leading to $(\alpha-1/2)/d=(\alpha^\prime-1/2)/d^\prime$).
As an example, if a one-dimensional system ($d^\prime=1$) should \anfz{imitate} the superradiant state 
from three-dimensional free space ($d=3$, $\alpha=1$), the electromagnetic environment
would have to be \anfz{engineered} \cite{kimble15} such that $\alpha^\prime=2/3$.
Similarly, if an extended sample should realise small-volume superradiance, transformation (\ref{eq:chitransform})
must reproduce the transformation of a Dicke system, 
which is simply 
$\chi_{\mathrm{Dicke}} \rightarrow \tilde{\chi}_{\mathrm{Dicke}}$
with ${\chi}_{\mathrm{Dicke}} = {N}$ and $\tilde{\chi}_{\mathrm{Dicke}} = \tilde{N} = f_N {\chi}_{\mathrm{Dicke}}$.
Hence, $\alpha=(1-d)/2 \geq 0$, which reveals that only extended samples in one dimension 
($d=1$, otherwise we would have $\alpha < 0$) can behave \anfz{Dicke-like}.

In conclusion, we have studied single-photon superradiance in extended media, 
and showed how superradiance can be engineered in such a way that an artificial optical transition with tunable decay rate and level shift is realised. 
This result provides the basic building block for a systematic approach towards engineering advanced artificial quantum systems via superradiance by design.
A promising avenue for future studies is the extension of our work to coupled 
sub ensembles with the goal to design artificial multi-level atoms \cite{roehlsb12}.

\section*{Methods}
For an extended lattice, the plane wave ansatz
$\varphi_{\vv{r}} = (1/\sqrt{N}) \ee^{\ii \vv{k} \vv{r}}$ 
for eq.~(\ref{eq:eigenproblem})
yields the eigenstates' decay rates~$\Gamma_{\vv{k}}=-2 \mathrm{Im}(E_{\vv{k}})$ 
and frequency shifts~$\Delta_{\vv{k}}=\mathrm{Re}(E_{\vv{k}}) - \omega_0$
as
\begin{eqnarray}
 \label{eq:eigenrates}
 \Gamma_{\vv{k}} &=& \gamma_{0} + \mathrm{Re}[\mathcal{I}_d(\vv{k})] \mathcomma
 \\
 \label{eq:eigenshifts}
 \Delta_{\vv{k}} &=& \delta \omega_{0} + \frac{1}{2} \mathrm{Im}[\mathcal{I}_d(\vv{k})] \mathcomma
 \\
 \label{eq:Iterm}
 \mathcal{I}_d(\vv{k}) &=&
 \sum_{\vv{r}}{}^{'}~
  V_{r}~ \ee^{- \ii \vv{k} \vv{r}}
 \mathperiod
\end{eqnarray}
Here, 
$\vv{r}=(x_1,\dots,x_d)^{T}$ denotes a $d$-dimensional lattice vector with components
$x_i=a n_i$, $i=1,\dots,d$, $n_i=-\sqrt[d]{N}/2+1,\dots,\sqrt[d]{N}/2$, and $\sqrt[d]{N}$ is even.
Likewise, $\vv{k}=(k_1,\dots,k_d)^T$ is the wavevector of the collective atomic excitation.
The sum runs over all combinations of $\{ n_i \}$ except $n_1=\dots=n_d=0$
and the couplings depend on the distance 
$r=\abssmall{\vv{r}}=a \sqrt{n_1^2 + \dots + n_d^2}$ between atoms.
We assume the atomic dipole moments to be 
uniformly aligned along the $x_3$ axis (\eg, by applying a weak magnetic field).
Thus, for $d=1,2$ the distance vector $\vv{r}$ (in the $x_1$-$x_2$ plane) is perpendicular to the 
dipole moments, and for $d=3$ we have to take into account the polar angle 
$\theta=\arccos( x_3 / r)$.
Furthermore, we make use of the assumptions $N \gg 1$ (many atoms) and 
$k_0 a > 1$ (extended sample).
The decay rate eq.~(\ref{eq:eigenrates}) can also be rewritten in terms of 
the enhancement factor
\begin{equation}
 \label{eq:chi}
 \chi_{\vv{k}} \equiv \frac{\Gamma_{\vv{k}}}{\gamma_0} = 1 + \frac{\mathrm{Re}[\mathcal{I}_d(\vv{k})]}{\gamma_0}
 \mathperiod
\end{equation}

To arrive at the final expressions eqs.~(\ref{eq:enhfac})-(\ref{eq:shiftsfiniteM}),
we further manipulate eqs.~(\ref{eq:eigenrates})-(\ref{eq:chi})
as follows.
In this paper, 
we focus on the system's eigenstates 
and---to keep the analysis general---do not consider geometric details 
or questions of how to excite and probe the system since 
such details vary from experiment to experiment.
Around $k\equiv\abssmall{\vv{k}}=k_0$, 
we can utilise a continuum formulation,
rewrite the lattice sums in eq.~(\ref{eq:Iterm}) into an integral, 
and perform the angular integration for couplings of type (\ref{eq:gencoupling}) 
(see supplementary information for further technical details of the calculation), 
leading to
\begin{eqnarray}
 \label{eq:radialint}
 \frac{\mathcal{I}_d}{\gamma_0} &=& 
 \frac{2 c_d}{\left( k_0 a\right)^d} 
%  \cdot \left(\frac{k}{k_0}\right)^{\frac{1}{2}(d-1)}
 \cdot \left(\frac{k_0}{k}\right)^{\frac{1}{2}(d-1)}
 \cdot A_d
 \cdot \mathcal{J}_d(k)
 \mathcomma \\
 \label{eq:radialintJ}
 \mathcal{J}_d(k) &=&
 \int \limits_{k_0 a}^{k_0 a b_d \sqrt[d]{N}} \mathrm{d}\eta~\ee^{\epsilon \ii \eta} {g}_d(k k_0^{-1} \eta)~\eta^{\beta}
 \mathcomma
 \\
 \label{eq:betadef}
 \beta &\equiv& \frac{d-1}{2} - \alpha 
 \mathperiod
\end{eqnarray}
The dimension-dependent quantities $A_d$, $b_d$, $c_d$, and $g_d$ are listed in table~\ref{tab:dimexpr}
(for instance, $A_d$ is real for $d=1,2$ and purely imaginary for $d=3$).
Note that
the factor $\exp(\pm \ii k_0 r)$ from eq.~(\ref{eq:gencoupling}) 
in the eigenproblem (\ref{eq:eigenproblem}) can be understood
as a radial translation in wavenumber space. 
In the shifted frame, a long-wavelength limit 
of the collective atomic excitation
(which can be accounted for by a continuum description) 
corresponds to $k \rightarrow k_0$.
This continuum formulation is applicable in the range
$\abssmall{k-k_0} \lesssim \pi / a b_d \sqrt[d]{N}$.
Further,
in eq.~(\ref{eq:gencoupling}), we have not included exponential damping
of the form $\exp(-k_0 r/\ell)$, where $\ell$ denotes a dimensionless absorption 
length that, for instance, empirically accounts for material imperfections.
Such a damping factor in the integral in eq.~(\ref{eq:radialintJ}) 
would lead to 
a broadening and modification of the $k=k_0$-criterion for maximal superradiance,
going beyond the scope of this paper.
Details on the calculation of the integrals in eq.~(\ref{eq:radialintJ}) can be found in the supplementary material.

The maximum enhancement factor (\ref{eq:enhfac})
can be cast into the equivalent forms 
($\mathcal{V}=Na^d$ denotes the sample volume, 
$\rho=N/\mathcal{V}$ is the number density, 
and $\chi_{\mathrm{max}}-1\simeq \chi_{\mathrm{max}}$ since $N \gg 1$)
\begin{align}
\label{eq:chiformuls1}
 \frac{\chi_{\mathrm{max}}}{{L}_d(\alpha)}  	&=  
		      \left( \frac{\lambda_0}{\sqrt[d]{\mathcal{V}}} \right)^{\frac{1}{2}(d-1)+\alpha} \cdot N	& (N,\mathcal{V})& \\[2ex]
  \label{eq:chiformuls2}
	~	&=   \left( \frac{\lambda_0}{\sqrt[d]{\mathcal{V}}} \right)^{\frac{1}{2}(d-1)+\alpha} \cdot \mathcal{V} \cdot \rho & (\mathcal{V},\rho)& \\[2ex]
  \label{eq:chiformuls3}
	~	&=   \left( {\lambda_0 \sqrt[d]{\rho}} 		    \right)^{\frac{1}{2}(d-1)+\alpha} \cdot \sqrt[d]{N}^{\frac{1}{2}(d+1)-\alpha} & (N,\rho)&\mathcomma \\[2ex]
  \label{eq:chiformuls4}
   {L}_d(\alpha) &\equiv \frac{2 \left(b_d\right)^{\frac{1}{2}(d+1)-\alpha} c_d}{ d+1 - 2\alpha } 
      \cdot \frac{\abs{A_d}}{\left(2\pi\right)^{\frac{1}{2}(d-1)+\alpha}} \mathperiod  & ~& 
\end{align}
Which formulation to choose from eqs.~(\ref{eq:chiformuls1})-(\ref{eq:chiformuls3}) depends on 
which quantities can be controlled in an experiment.

If for small volumes the length scale set by the inter-atomic distance $a$ 
is effectively eliminated from the single-excitation eigenproblem (\ref{eq:eigenproblem})
(possibly neglecting divergent contributions to the inter-atomic coupling \cite{haroche82,garraway11}),
all atoms couple to each other with equal strength $V_0$.
The resulting equation $E-\omega_0=-(\ii/2) V_0 \sum_{j=1}^N \varphi_{\vv{r}_j} / \varphi_{\vv{r}_i}$
(which must hold for all $\vv{r}_i$) yields a maximal decay rate for a spatially constant wavefunction  
with equal relative phase between all atom pairs, representing the maximally symmetric Dicke state.
For this state, 
$\Gamma = -2 \mathrm{Im}(E) = N \gamma_0$
and 
$\Delta = \mathrm{Re}(E) - \omega_0 = N \delta\omega_0$.

%\clearpage

\section*{Author contributions}
P.L. initiated the project, performed the calculations, and, together with J.E., interpreted the results.
P.L. and J.E. wrote the manuscript.
All authors contributed to the development of ideas, critical discussions, and the preparation of the manuscript.
C.K. and J.E. guided the project.

\section*{Competing financial interests}
The authors declare no competing financial interests.

\clearpage

\section*{Supplementary information}
In this supplementary information, we provide technical details on how to rewrite the lattice sums 
into an integral and on how to ultimately perform the integration.

% \author{Paolo~Longo}
% % \email{paolo.longo@mpi-hd.mpg.de}
% \author{Christoph H. Keitel}%
% \author{J\"org~Evers}%
% \affiliation{Max Planck Institute for Nuclear Physics, Saupfercheckweg 1, 69117 Heidelberg, Germany}
% 
% 
% % \date{\today}
% 
% 
% \begin{abstract}
% {
% In this supplementary information, we provide technical details on how to rewrite the lattice sums 
% into an integral and on how to ultimately perform the integration.
% }
% \end{abstract}
% 
% 
% 
% 
% \maketitle

\section{$\mathcal{I}_d(\vv{k})$}
The quantity
\begin{equation}
  \mathcal{I}_d(\vv{k}) =
 \sum_{n_1}{}^{'} \dots \sum_{n_d}{}^{'}
  V_{a \sqrt{n_1^2 + \dots + n_d^2} } \ee^{- \ii k_1 a n_1} \dots \ee^{- \ii k_d a n_d}
\mathcomma
\end{equation}
where the sums run over all combinations of $\{ n_i \}$ except $n_1=\dots=n_d=0$,
can be rewritten into an integral 
\begin{equation}
 \mathcal{I}_d(\vv{k}) \rightarrow 
\int \frac{\mathrm{d}^dx}{a^d} V(r,\theta) \prod_{j=1}^d \ee^{- k_j x_j}
\mathperiod
\end{equation}
For $d=1,2$, $V(r,\theta)=f_r$, whereas for $d=3$,
$V(r,\theta) = \sin^2\theta~f_r$ ($f_r \equiv A_d \exp(\epsilon \ii k_0 r)/(k_0 r)^\alpha$).
Here, $\mathrm{d}^dx$ signifies the $d$ dimensional infinitesimal volume element
and the integration is over all space except for a region with radius $a$ around the origin
($\mathrm{d}^1 x=\mathrm{d} x$, 
$\mathrm{d}^2 x= r \mathrm{d} r \mathrm{d} \varphi$,
$\mathrm{d}^3 x= r^2 \sin\theta \mathrm{d} r \mathrm{d} \varphi \mathrm{d} \theta$).

Explicitly, for $d=1$,
\begin{eqnarray}
 \mathcal{I}_1 &=& \int_{-\frac{Na}{2}}^{\frac{Na}{2}} \frac{\mathrm{d}x}{a} \Theta(\abs{x}-a)
  f_{\abs{x}} \ee^{-\ii kx} \\ \nonumber
 &=&
 \int_a^{\frac{Na}{2}} \frac{\mathrm{d}x}{a}  f_{\abs{x}} 2 \cos(kx) \\ \nonumber
&=&
 2 \int_a^{\frac{Na}{2}} \frac{\mathrm{d}r}{a}  f_{r} \cos(kr)
 \mathcomma
\end{eqnarray}
where $k=\abssmall{\vv{k}}$ and $\Theta(\cdot)$
signifies the Heaviside step function.
For $d=2$, 
\begin{eqnarray}
 \mathcal{I}_2 &=&
 \int_a^{N^\prime_2 a} \int_0^{2\pi} \frac{r\mathrm{d}r\mathrm{d}\varphi}{a^2}
 f_r \ee^{-\ii k_1 r \cos\varphi} \ee^{-\ii k_2 r \sin\varphi}
 \\ \nonumber
 &=& \frac{2\pi}{a^2} \int_a^{N^\prime_2 a} \mathrm{d}r~ r f_r J_0(kr)
\mathcomma
\end{eqnarray}
where $N^\prime_2$ is chosen such that the integration area covers $N$ atoms,
\ie, $\pi (N^\prime_2)^2 = N$.
For $d=3$, 
\begin{eqnarray}
 \mathcal{I}_3 &=&
 \int_a^{N^\prime_3 a} \frac{r^2 \mathrm{d}r}{a^3}
 \int_0^\pi \mathrm{d}\theta \sin\theta
 \int_0^{2\pi} \mathrm{d}\varphi \underbrace{V(r,\theta)}_{=\sin^2\theta~ f_r} 
 \\ \nonumber
 && ~~~\times
 \ee^{-\ii k_1 r \sin\theta \cos\varphi}
 \ee^{-\ii k_2 r \sin\theta \sin\varphi}
 \ee^{-\ii k_3 r \cos\theta}
 \\ \nonumber
 &=&
 2 \pi
 \int_a^{N^\prime_3 a} \frac{r^2 \mathrm{d}r}{a^3} f_r
 \int_0^\pi \mathrm{d}\theta \sin^3\theta
 \ee^{-\ii k_3 r \cos\theta}
 \\ \nonumber
 && ~~~~~~~~~~\times
 J_0\left(\sin\theta \sqrt{k_1^2+k_2^2}~r\right)
 \mathcomma
\end{eqnarray}
where $(4\pi/3) (N^\prime_3)^3=N$, $J_0(\cdot)$ signifies the zeroth-order
Bessel function of first kind, and $\mathrm{sinc}(x)=\sin(x)/x$.
The integration over $\theta$ can be done as follows.
Upon definining 
$I_m \equiv \int_{0}^\pi \mathrm{d}\theta \sin^m\theta \exp(-\ii k_{\perp} r \cos\theta) J_0(k_{\parallel} r \sin\theta)$, 
$k_{\perp} \equiv k_3$, and $k_{\parallel}\equiv \sqrt{k_1^2+k_2^2}$, we have the relation
\begin{eqnarray}
 I_3 &=& I_1 - \int_{0}^\pi \mathrm{d}\theta \cos^2\theta \sin\theta \ee^{-\ii k_{\perp} r \cos\theta} J_0(k_{\parallel} r \sin\theta)
 \\ \nonumber
 &=& \left( 1 + \frac{1}{r^2} \frac{\partial^2}{\partial k_{\perp}^2} \right) I_1
 \mathperiod
\end{eqnarray}
To simplify the integration needed for $I_1$, we can choose a coordinate system in which 
either $k_{\perp}=0$ or $k_{\parallel}=0$ 
(it can be shown that $I_1$ does not depend on the orientation of $\vv{k}$),
yielding
\begin{equation}
 I_1 = 2~ \mathrm{sinc} \left( \sqrt{k_{\perp}^2 + k_{\parallel}^2}  ~r \right)
 \mathperiod
\end{equation}
Finally, 
$I_3 = \sin^2\vartheta \cdot 2 \mathrm{sinc}(kr) + \mathcal{O}[(kr)^{-2}]$
and therefore
\begin{equation}
 \mathcal{I}_3 = 
  \frac{4 \pi}{a^3} \sin^2\vartheta \int_a^{N^\prime_3 a} \mathrm{d}r~r^2 f_r 
 \mathrm{sinc}(kr)
 \mathcomma
\end{equation}
where 
$\vartheta$ denotes the angle between the eigenstate's wavevector $\vv{k}$ and the $z$ axis.
Here, we have only taken into account the asymptotic leading order term (with respect to $kr$).
Other terms can be accounted for by means of different coefficients $\alpha$ (see main text).
Introducing the abbreviations used in the paper and performing a variable substituion, we finally 
arrive at the integrals $\mathcal{J}_d(k)$ (Eq.~(14) in the main text).

\section{$\mathcal{J}_d(k)$}
We now proceed with the radial integration
\begin{eqnarray}
 \mathcal{J}_d(k) &=&
 \int \limits_{k_0 a}^{k_0 a b_d \sqrt[d]{N}} \mathrm{d}\eta~\ee^{\pm \ii \eta} {g}_d(k k_0^{-1} \eta)~\eta^{\beta}
 \mathperiod
\end{eqnarray}
Note that for $d=2$ the integration kernel is actually given by $J_0(kk_0^{-1}\eta)$.
However, already at the lower integration limit, we can use the asymptotic form 
$J_0(ka) \approx \sqrt{2/\pi} \cos(ka-\pi/4)/\sqrt{ka}$ since in an extended sample 
$k_0 a>1$ and only the wavenumbers $k$ around $k_0$ are relevant. 
Furthermore, by an additional substituion of the integration variable, we 
shift the $\pi/4$ shift to the argument to the exponential (which we can account for by means of appropriate prefactors),
$\eta+\pi/4 \approx \eta$, and the integration limits can also approximately remain unchanged.

The possible combinations in the integrand we need to consider are 
\begin{eqnarray}
 {J}_{\mathrm{cc}} &\equiv&
 \int \limits_{k_0 a}^{k_0 a b_d \sqrt[d]{N}} \mathrm{d}\eta~\cos(\eta) \cos(k k_0^{-1} \eta)~\eta^{\beta}
 \mathcomma
 \\
 {J}_{\mathrm{sc}} &\equiv&
 \int \limits_{k_0 a}^{k_0 a b_d \sqrt[d]{N}} \mathrm{d}\eta~\sin(\eta) \cos(k k_0^{-1} \eta)~\eta^{\beta}
 \mathcomma
 \\
 {J}_{\mathrm{cs}} &\equiv&
 \int \limits_{k_0 a}^{k_0 a b_d \sqrt[d]{N}} \mathrm{d}\eta~\cos(\eta) \sin(k k_0^{-1} \eta)~\eta^{\beta}
 \mathcomma
 \\
 {J}_{\mathrm{ss}} &\equiv&
 \int \limits_{k_0 a}^{k_0 a b_d \sqrt[d]{N}} \mathrm{d}\eta~\sin(\eta) \sin(k k_0^{-1} \eta)~\eta^{\beta}
 \mathperiod
\end{eqnarray}

\subsection{${J}_{\mathrm{cc}}$}
Utilizing a computer algebra system, we find that
\begin{eqnarray}
 J_{cc} &=& \frac{\ii^{\beta +1}}{4} \cdot
    \Bigg{[}
    \left( \frac{r_-}{k_0} \right)^{-1-\beta} \left( \mathrm{sgn}(r_-) \right)^{-2 \beta}
  \\ \nonumber
  && ~~~~~~~~~~ \times
    \Big{(} \Gamma(1+\beta,-\ii r_- a) 
  \\ \nonumber && ~~~~~~~~~~~~~~~~ - \Gamma(1+\beta, -\ii r_- a N^\prime_d) 
  \\ \nonumber && ~~~~~~~~~~~~~~~~ - (-1)^\beta \Gamma(1+\beta, \ii r_- a) 
  \\ \nonumber && ~~~~~~~~~~~~~~~~ + (-1)^\beta \Gamma(1+\beta, \ii r_- a N^\prime_d) 
    \Big{)}
    \\ \nonumber
    && ~~~~~ ~+~ 
    \left( \frac{r_+}{k_0} \right)^{-1-\beta} 
    \\ \nonumber
    && ~~~~~~~~~~ \times
    \Big{(} \Gamma(1+\beta,-\ii r_+ a) 
      \\ \nonumber && ~~~~~~~~~~~~~~~~ - \Gamma(1+\beta, -\ii r_+ a N^\prime_d) 
      \\ \nonumber && ~~~~~~~~~~~~~~~~ - (-1)^\beta \Gamma(1+\beta, \ii r_+ a) 
      \\ \nonumber && ~~~~~~~~~~~~~~~~ + (-1)^\beta \Gamma(1+\beta, \ii r_+ a N^\prime_d) 
    \Big{)}
    \Bigg{]}
 \mathcomma
\end{eqnarray}
where $r_{\pm} \equiv k \pm k_0$ and 
$\Gamma(a,z) \equiv \int_z^\infty \mathrm{d}t~t^{a-1} \ee^{-t}$ 
signifies the incomplete Gamma function.
The asymptotic form for $N^\prime_d \gg 1$ and $r_- \neq 0$
reads
\begin{eqnarray}
 J_{cc}(k \neq k_0) &\simeq& \frac{\ii^{\beta +1}}{4} \cdot
    \Bigg{[}
    \left( \frac{r_-}{k_0} \right)^{-1-\beta} \left( \mathrm{sgn}(r_-) \right)^{-2 \beta}
  \\ \nonumber
  && ~~~~~~~~~~ \times
    \Big{(} \Gamma(1+\beta,-\ii r_- a) 
  \\ \nonumber && ~~~~~~~~~~~~~~~~ - (-1)^\beta \Gamma(1+\beta, \ii r_- a)
  \\ \nonumber && ~~~~~~~~
	  - 2 \ii (-1)^\beta (\ii r_- a N^\prime_d)^\beta \sin(r_- a N^\prime_d)
    \Big{)}
    \\ \nonumber
    && ~~~~~ ~+~ 
    \left( \frac{r_+}{k_0} \right)^{-1-\beta} 
    \\ \nonumber
    && ~~~~~~~~~~ \times
    \Big{(} \Gamma(1+\beta,-\ii r_+ a) 
  \\ \nonumber && ~~~~~~~~~~~~~~~~ - (-1)^\beta \Gamma(1+\beta, \ii r_+ a)
  \\ \nonumber && ~~~~~~~~ - 2 \ii (-1)^\beta (\ii r_+ a N^\prime_d)^\beta \sin(r_+ a N^\prime_d)
    \Big{)}
 \mathperiod
\end{eqnarray}
The dominant terms for $N^\prime_d \gg 1$ in this expression are $\propto (N^\prime_d)^\beta$
if $\beta>0$.

For $k \rightarrow k_0$ ($N^\prime_d = b_d \sqrt[d]{N} \gg 1$),
we arrive at
\begin{eqnarray}
 J_{cc}(k \rightarrow k_0) &\simeq&  \frac{ 1 }{2(1+\beta)} 
		(k_0 a)^{\beta+1} \left( N^\prime_d \right)^{\beta+1}
  \\ 
  &=&
  \frac{(b_d)^{\beta+1}}{2(1+\beta)} 
		(k_0 a)^{\beta+1}  N^{\frac{\beta+1}{d}}
 \mathperiod
\end{eqnarray}
Here, the dominant terms for $N^\prime_d \gg 1$
are $\propto (N^\prime_d)^{\beta+1}$ if $\beta > -1$.

For the case $\beta=0$, the explicit expression for the integral reads
\begin{eqnarray}
 \nonumber
 J_{cc} &\overset{\beta=0}{=}& \frac{k_0 a}{2} 
  \Big{[} N^\prime_d \left( \mathrm{sinc}(r_- a N^\prime_d) + \mathrm{sinc}(r_+ a N^\prime_d)  \right) 
  \\ 
 && ~~~~~~~~
			  -\mathrm{sinc}(r_+ a) - \mathrm{sinc}(r_- a)
 \Big{]}
 \\ \nonumber
 &\overset{\abssmall{r_-} a \ll 1,~N^\prime_d \gg 1}{\simeq}& 
 \frac{k_0 a}{2} N^\prime_d \mathrm{sinc}(r_- a N^\prime_d)
 \\ \nonumber
	& \overset{k \rightarrow k_0}{\rightarrow} & \frac{k_0 a}{2} N^\prime_d  
	= \frac{k_0 a}{2} b_d N^{\frac{1}{d}}
	\mathperiod
\end{eqnarray}
In the second step, we focus on wavenumbers $k$ around $k_0$ (\ie, small $\abssmall{r_-}a$,
for which the terms with $r_+$ are negligible).

\subsection{${J}_{\mathrm{sc}}$}
Similarly, 
\begin{eqnarray}
 J_{sc} &=& \frac{\ii^{\beta +2}}{4} \cdot
    \Bigg{[}
    \left( \frac{r_-}{k_0} \right)^{-1-\beta} \left( \mathrm{sgn}(r_-) \right)^{-2 \beta} 
\\ \nonumber
  && ~~~~~~~~~~ \times
    \Big{(} - \Gamma(1+\beta,-\ii r_- a) 
\\ \nonumber && ~~~~~~~~~~~~~~~~+ \Gamma(1+\beta, -\ii r_- a N^\prime_d) 
\\ \nonumber && ~~~~~~~~~~~~~~~~  - (-1)^\beta \Gamma(1+\beta, \ii r_- a) 
\\ \nonumber && ~~~~~~~~~~~~~~~~+ (-1)^\beta \Gamma(1+\beta, \ii r_- a N^\prime_d) 
    \Big{)}
    \\ \nonumber
    && ~~~~~ ~+~ 
    \left( \frac{r_+}{k_0} \right)^{-1-\beta} 
    \Big{(} \Gamma(1+\beta,-\ii r_+ a) 
\\ \nonumber && ~~~~~~~~~~~~~~~~ - \Gamma(1+\beta, -\ii r_+ a N^\prime_d) 
\\ \nonumber && ~~~~~~~~~~~~~~~~  + (-1)^\beta \Gamma(1+\beta, \ii r_+ a) 
\\ \nonumber && ~~~~~~~~~~~~~~~~ - (-1)^\beta \Gamma(1+\beta, \ii r_+ a N^\prime_d) 
    \Big{)}
    \Bigg{]}
 \mathperiod
\end{eqnarray}
The asymptotic form for $N_d^\prime \gg 1$ but $r_- \neq 0$
reads
\begin{eqnarray}
 J_{sc}(k \neq k_0) &\simeq& \frac{\ii^{\beta +2}}{4} \cdot
    \Bigg{[}
    \left( \frac{r_-}{k_0} \right)^{-1-\beta} \left( \mathrm{sgn}(r_-) \right)^{-2 \beta}
\\ \nonumber
  && ~~~~~~~~~~ \times
    \Big{(} -\Gamma(1+\beta,-\ii r_- a) 
\\ \nonumber && ~~~~~~~~~~~~~~~~ - (-1)^\beta \Gamma(1+\beta, \ii r_- a)
\\ \nonumber && ~~~~~~~~
	  + 2 (-1)^\beta (\ii r_- a N^\prime_d)^\beta \cos(r_- a N^\prime_d)
    \Big{)}
    \\ \nonumber
    && ~~~~~ ~+~ 
    \left( \frac{r_+}{k_0} \right)^{-1-\beta} 
    \Big{(} \Gamma(1+\beta,-\ii r_+ a) 
\\ \nonumber && ~~~~~~~~~~~~~~~~ - (-1)^\beta \Gamma(1+\beta, \ii r_+ a)
\\ \nonumber && ~~~~~~~~
	  - 2 (-1)^\beta (\ii r_+ a N^\prime_d)^\beta \cos(r_+ a N^\prime_d)
    \Big{)} \Bigg{]}
 \mathperiod
\end{eqnarray}

In contrast to the integral $J_{cc}$, the limit $k \rightarrow k_0$
does not yield a scaling $\propto (N^\prime_d)^{\beta + 1}$.
The case $\beta = 0$ reads 
\begin{eqnarray}
 J_{sc} &\overset{\beta=0}{=}& \frac{k_0 a}{2} 
	  \Big[ 
	    \frac{\cos(r_- a N^\prime_d)}{r_- a}  - \frac{\cos(r_+ a N^\prime_d)}{r_+ a}
\\ \nonumber 
&& ~~~~~~~~
	    - \frac{\cos(r_- a)}{r_- a} + \frac{\cos(r_+ a)}{r_+ a}
	  \Big]
	  \\ \nonumber
  &\overset{\abssmall{r_-} a \ll 1}{\simeq}& 
  \frac{k_0 a}{2} \cdot 
	    \frac{\cos(r_- a N^\prime_d)-1}{r_- a} 
 \mathperiod
\end{eqnarray}

\subsection{${J}_{\mathrm{cs}}$}
We can rewrite
\begin{eqnarray}
 {J}_{\mathrm{cs}} &=&
 \int \limits_{k_0 a}^{k_0 a b_d \sqrt[d]{N}} \mathrm{d}\eta~\cos(\eta) \sin(k k_0^{-1} \eta)~\eta^{\beta}
 \\ \nonumber
 &=& 
 \left( \frac{k_0}{k} \right)^{\beta+1}
 \int \limits_{k a}^{k a b_d \sqrt[d]{N}} \mathrm{d}\eta~\sin(\eta) \cos(k_0 k^{-1} \eta)~\eta^{\beta}
 \mathperiod
\end{eqnarray}
This is in essence just the integral $J_{\mathrm{sc}}$ with $k$ and $k_0$ interchanged.
In particular, for $\beta=0$, 
${J}_{\mathrm{cs}} =  - J_{\mathrm{sc}}.$

\subsection{${J}_{\mathrm{ss}}$}
For $k \rightarrow k_0$, we can write
\begin{eqnarray}
 {J}_{\mathrm{ss}}(k\rightarrow k_0) &=&
 \int \limits_{k_0 a}^{k_0 a b_d \sqrt[d]{N}} \mathrm{d}\eta~(1-\cos^2\eta)~\eta^{\beta+1}
 \\ \nonumber
 &=&
 \underbrace{(k_0 a b_d \sqrt[d]{N})^{\beta+1}}_{2 J_{\mathrm{cc}}(k \rightarrow k_0)} - (k_0 a)^{\beta+1} - J_{\mathrm{cc}}(k\rightarrow k_0)
\\ \nonumber
 &\overset{N \gg 1}{\simeq}&
J_{\mathrm{cc}}(k \rightarrow k_0)
\mathperiod
\end{eqnarray}

For $\beta=0$, the explicit expression reads
\begin{eqnarray}
 \nonumber
 J_{ss} &\overset{\beta=0}{=}& \frac{k_0 a}{2} 
  \Big{[} N^\prime_d \left( \mathrm{sinc}(r_- a N^\prime_d) - \mathrm{sinc}(r_+ a N^\prime_d)  \right) 
  \\ 
 && ~~~~~~~~
			  +\mathrm{sinc}(r_+ a) - \mathrm{sinc}(r_- a)
 \Big{]}
 \\ \nonumber
 &\overset{\abssmall{r_-} a \ll 1,~N^\prime_d \gg 1}{\simeq}& 
 \frac{k_0 a}{2} N^\prime_d \mathrm{sinc}(r_- a N^\prime_d)
 \mathperiod
\end{eqnarray}

\end{document}